\input phyzzx    

\twelvepoint
%\font\small =cmr10 scaled 680
%\font\bb=msbm10 scaled 1200

%%%%%%%%%%%%%%%%%%%%%%Some Definitions%%%%%%%%%%%%%%%%%%%

\def\a{\alpha'}

\def\hX{{\hat X}}
\def\hu{{\hat u}}

%%%%%%%%%%%%%%%%%%%%%%%%%%%%References%%%%%%%%%%%%%%%%%%%

\REF\Haagensen{P. Haagensen, Phys. Lett. {\bf B382} (1996) 356, hep-th/9604136}
\REF\Douglas{M.R. Douglas, {\it Gauge Fields and D-branes}, hep-th/9604198;
M.R. Douglas, D. Kabat, P. Pouliot, and S.H. Shenker, 
{\it D-branes and Short Distances in String Theory},  hep-th/9608024}
\REF\HPT{C.M. Hull, G. Papadopoulos and P.K. Townsend, Phys. Lett. {\bf B316} 
(1993) 187, hep-th/9307013; 
G. Papadopoulos and P.K. Townsend, Class. Quant. 
Grav. {\bf 11} (1994) 515, hep-th/9307066; 
G. Papadopoulos and P.K. Townsend, Class. Quant. 
Grav. {\bf 11} (1994) 2180, hep-th/9406015}
\REF\AF{L. Alvarez-Gaum{\'e} and D.Z. Freedman, Comm. Math. Phys. {\bf 91} 
(1983) 87}
\REF\Hassan{S.F. Hassan, Nucl. Phys. {\bf B460} (1996) 362,  hep-th/9504148}
\REF\Sfetsos{K. Sfetsos, in
{\it Gauge Theories, Applied Supersymmetry and Quantum Gravity},
Leuven University Press, Leuven, 1996, hep-th/9510103}
\REF\HHT{J.C. Henty, C.M. Hull and P.K. Townsend, 
Phys. Lett. {\bf B185} (1987) 73}
\REF\Lambert{N.D. Lambert, Nucl. Phys. {\bf B469} (1996) 68, hep-th/9510130}
\REF\AT{E.R.C. Abraham and P.K. Townsend, Phys. Lett. {\bf B291} (1992) 85}
\REF\CHS{C. Callan, J. Harvey and A. Strominger, Nucl Phys. {\bf B359} (1991)
 611}

%%%%%%%%%%%%%%%%%%%%%%%%%%%%Title Page%%%%%%%%%%%%%%%%%%%
\pubnum={DAMTP-R/97/13\cr KCL-TH-97-25\cr hep-th/9703143}
\date{May 1997}

\titlepage

\title{\bf Potentials and T-Duality}

\bigskip

\centerline{S. F. Hewson\foot{sfh10@damtp.cam.ac.uk}}

\address{DAMTP\break
         Silver Street, Cambridge\break
         England\break
         CB3 9EW}

\bigskip
\centerline{and}
\bigskip

\centerline{N.D. Lambert\foot{lambert@mth.kcl.ac.uk}}

\address{Department of Mathematics\break
         King's College, London\break
         England\break
         WC2R 2LS}

\vfil

\abstract

We discuss the application of T-duality to 
massive supersymmetric sigma models. In particular 
$(1,1)$ supersymmetric models with off-shell central charges
reveal  an interesting structure. The T-duality transformations of the
BPS states of these theories
are also discussed and an explicit example of Q-kinks is given.

\endpage

%%%%%%%%%%%%%%%%%%%%%%%%%%%%%%%%Chapter One%%%%%%%%%%%%%%%

\chapter{Introduction}
 T-duality is 
a central tool in the the study of superstring vacua. While T-duality
of massless sigma models has been studied at  
length the, T-duality of massive sigma models has received little
attention. This is  
perhaps 
because the presence of the potential terms violates conformal
invariance, thus obscuring 
 the connection with perturbative string theory and conformal field 
theory. However,  at the level of the sigma model  action, 
T-duality 
is simply a Gaussian integral over a Killing direction and thus  the
duality may in this sense be applied to a massive sigma model.  
Recent advances in non-perturbative string theory have shown 
that in the  presence of 
D-branes, string  worldsheets  do  
possess massive terms 
[\Douglas];   one may  anticipate that the complete role of worldsheet 
potentials
has yet to be realised. It is therefore of some interest to
understand the implications of T-duality for such theories. Much of
the discussion is then readily extended to other  $p$-brane actions
with potentials. 
 
Consider the following bosonic sigma model action 
$$
S = {1\over 2\pi\a}\int d^2\sigma\ \left\{
(g_{IJ}\eta^{\mu\nu}+b_{IJ}\epsilon^{\mu\nu})
\partial_{\mu}X^I\partial_{\nu} X^J - V(X)\right\} \ ,
\eqn\introa
$$
where $I,J = 0,1,2,...,(D-1)$ and $V$ 
is a potential term. Let us suppose that the
target space has a $U(1)$ isometry associated with the direction
$X^0$, with the other coordinates labelled by $X^i$.  
In this simple bosonic model the potential transforms trivially under
the duality, since it is independent of $X^0$.  Furthermore, if we
view \introa\ as the bosonic  
sector
of a supersymmetric theory then we again find that the potential
transforms trivially by using 
$(1,1)$ superfields. However,  it is precisely in the cases 
for which the target space possesses
a Killing vector $k = \partial/\partial X^0$  that there is another
form of the potential compatible with $(1,1)$ supersymmetry [\AF,\HPT]. 
This potential 
can not be written using (standard) $(1,1)$ superfields because of the
appearance of off-shell central charges, and instead must  be
defined  as the length of the Killing vector $k$ 
$$
V ={1\over4}m^2 g_{IJ}k^I k^J 
= {1\over4}m^2g_{00} \ , 
\eqn\introb
$$
where $m$ is an arbitrary mass parameter. 
For the simplest
background with $g_{0i}=b_{0i}=0$,  duality inverts the metric
component $g_{00}$.  Naively following  the usual description of 
T-duality for
the massless sigma model,  the potential for the dual
background in such a case would  be  $V^{-1}$; hence it
would appear that  
supersymmetry could  be spontaneously broken in the dual
theory!   A related problem is that these theories
can possess BPS solitons interpolating between the different vacua of
the potential and these  must always exist in the
spectrum. The main motivation of this paper is to resolve these issues and 
examine the behaviour of the BPS states under T-duality.  We will 
see that the potential is always invariant under T-duality and that the 
BPS multiplets are also preserved. However, the dual potential has a different
form in
superspace and correspondingly different BPS solutions.

%%%%%%%%%%%%%%%%%%%%%%%%%%%%%%%%Chapter Two%%%%%%%%%%%%%%%

\chapter{The dual potential}

We initially consider supersymmetric sigma-models with  potentials
defined by the  
Killing vector 
$k=\partial/\partial X^0$. In such a case we may not express the
action in terms of  
$(1,1)$ superfields since the algebra possesses off-shell central charges 
[\HPT], and we lose 
manifest supersymmetry. In terms
of $(1,0)$ superfields the action is 
$$
S = -i\int d^2\sigma d\theta^+\ (g_{IJ}+b_{IJ})D_+\Phi^I\partial_=\Phi^J
+ ig_{IJ}\Psi^I_-\nabla^{(+)}_+\Psi^J + ims_I\Psi^I \ .
\eqn\six
$$
Here $\Phi^I=X^I + \theta^+\psi^I_+$,  $\Psi^I = \psi^I_- - \theta^+F^I$ 
and
$\nabla^{(+)}$ is
the covariant derivative with torsion. The action \six\ has manifest $(1,0)$
supersymmetry for any choice of the co-vector field $s_I$.\foot{There is a 
much more general class of $(1,0)$ 
supersymmetric sigma models than \six\  where $\Psi_-$ and $s$ 
lie in an arbitrary vector bundle over the target space
[\HPT]. It is not hard to
see from the discussion below that these fields, which include the mass terms,
transform trivially under T-duality unless the vector bundle is identified 
with the tangent bundle.} 
Elimination of  the auxiliary field by its equation of 
motion $2g_{IJ}F^J - ms_I = 0$ 
leads to the potential $V = {1\over4}m^2 g^{IJ}s_Is_J$.

Adapting the results of  [\Hassan] to $(1,0)$ supersymmetry
we find  the relation between the 
dual and original  superfields to be
$$\eqalign{
D_+{\hat \Phi}^0&=
-g_{00}D_+\Phi^0 - (g_{0i}+b_{0i})D_+\Phi^i \cr
D_+{\hat \Psi}^0_-&= 
-g_{00}D_+\Psi_-^0 - (g_{0i}+b_{0i})D_+\Psi_-^i\cr
\partial_={\hat \Phi}^0&= 
+g_{00}\partial_=\Phi^0 + (g_{0i}-b_{0i})\partial_=\Phi^i\cr
{\hat \Phi}^i&=\Phi^i \cr
{\hat \Psi}^i_-&= \Psi^i_-\ . \cr}
\eqn\sixb
$$
To dualise in such a way as to preserve manifest $(1,0)$
supersymmetry we must remove the dual auxiliary 
field by it's equation of motion in the dual theory. Equation 
\sixb\ implies that
$$
{\hat F}^0 = -g_{00}F^0 - (g_{0i}+b_{0i})F^i\,,\hskip1cm
{\hat F}^i = F^i \ .
\eqn\sevenb
$$
Now we note that this transformation is consistent with the equations of
motion in the dual model if and only  if
$$
{\hat s}_0=-{s_0\over g_{00}}\,,\hskip1cm
{\hat s}_i=s_i - ({g_{0i}\over g_{00}}+{b_{0i}\over g_{00}})s_0 \ .
\eqn\sevenc
$$
However, one can easily check that
${\hat V} = {\hat g}_{IJ}{\hat F}^I{\hat F}^J = g_{IJ}F^IF^J = V$. 
Thus the potential is invariant under T-duality even though the co-vector
is not. The supersymmetry is preserved and the vacua of the dual theory are
precisely the same as those in the  original theory.
We can also see from \sevenc\ that if $s_0=0$, as is the
case for models possessing a $(1,1)$ superspace form,  then both the 
potential and $s_I$ are invariant under T-duality.

We now consider the case when the action \six\  admits 
$(1,1)$ supersymmetry. The co-vector field must take the form 
$s_I = k_I-u_I$,  where
$\partial_{[J}u_{K]}=k^IH_{IJK}$ and $k^Iu_I=0$ [\HPT]. For our choice of
coordinates we find that 
$$
s_0 = g_{00} \,,\hskip1cm
s_i = g_{0i} + \left(b_{0i} - \partial_i\lambda\right)\ ,
\eqn\sevene
$$
where $\lambda$ is an arbitrary function of $X^i$, which may be
absorbed into $b_{0i}$, yielding the potential
$$
V = {1\over4}m^2\left(g_{00} + g^{ij}b_{0i}b_{0j}\right) \ . 
\eqn\eight
$$

To dualise this model we apply the
transformation \sevenc\ to obtain the  dual form of the co-vector field
$$
{\hat s}_0=-1\,,\hskip1cm
{\hat s}_i=0 \ .
\eqn\eightb
$$
It follows as before that ${\hat V}=V$. It is easy to see that this new form
also possess $(1,1)$ supersymmetry, only this time without off-shell central
charges. The obstruction to a full superspace expression is the fact
that the superspace potential ${\hat \Lambda}={\hat X}^0$ is only
locally defined in the target space, unlike the globally defined
derivative terms which appear in the equations of motion.  
Thus there are effectively two classes of $(1,1)$
supersymmetric models which do not possess a $(1,1)$ superspace form: those
defined in terms of Killing vectors,  and
those defined by topologically non-trivial closed forms. T-duality exchanges
these two classes and so in a sense exchanges  the isometry for 
the cohomology.

An important feature of the T-duality of a massless sigma model is that if
the $\beta$-functions of the original model vanish, then so do the 
$\beta$-functions of the dual model (they are in fact  
``covariant'' under T-duality
[\Haagensen]). A natural question to ask is whether or not the 
$\beta$-functions of the dual massive model are invariant under
duality, and also to investigate the effects of the massive terms on the beta
functions. One can easily see from dimensional analysis that the metric
and anti-symmetric tensor $\beta$-functions of a massive sigma model are
unaffected by the mass terms. Thus these must vanish in the dual model if
they vanish in the original one. So let us just consider the
massive $\beta$-function  here.
Following the result of [\HHT] and combining it with the
one loop divergence found in [\Lambert], whose notation we adopt, we
find that the contribution to the trace anomaly is 
$$
\beta_m = m\int d^2xd\theta^+\left\{
s_I - {1\over2}\alpha'
\left(\nabla^{(-)2}s_I - 2\partial^K\phi\nabla^{(-)}_Ks_I\right)
\right\}\Psi_-^I\ ,
\eqn\eleven
$$
where $\phi$ is the dilaton,  which enters through the Fradkin-Tseytlin term
in the action.
This integral consists of a classical contribution, caused by the
addition of the potential term to the conformally invariant massless sigma
model, and the order $\alpha'$ quantum corrections. 
We wish to investigate in which situations the order $\alpha'$ piece vanishes,
and also whether or not the form of the integral is invariant under
the T-duality procedure. 
In order to proceed, we restrict ourselves to the bosonic sector,
for which  $\Psi_-^I={1\over2}m\theta^+s^I$. The classical
contribution to \eleven  is simply the potential $V$, which is
invariant under T-duality. 
The order $\alpha'$ term is less simple: in general
it is
neither zero, nor  invariant under T-duality. In addition, 
contrasting  the the massless case, the
potential $\beta$-function is not covariant under  
T-duality. There are, however,
certain special classes of cases for which we can say something more
constructive: If $b_{0i}=0$ then we find that 
both the original and dual expressions vanish, and the anomaly is
entirely classical. Note that according to the torsion, the $b_{IJ}$ field is
only defined up to a total derivative; specifying $b_{0i}$ to be zero
is equivalent to making   a particular  choice for
$\lambda$ in the definition \sevene,  corresponding  to the
case for which the potential  is given by \introb.  This  includes
the cases of hyper-K\"{a}hler string backgrounds such as the 
Q-kink soliton sigma model, discussed later.  Similar  requirements on
$\lambda$ have  
been noted
previously in the context of one loop finiteness [\Lambert]. Finally,
since $b_{0i}\leftrightarrow g_{0i}$ under T-duality, $(1,0)$ 
the previous comments imply that the anomaly also vanishes for any
sigma model with $s_I$ of
the form \eightb  with vanishing metric  cross terms $g_{i0}$.

%%%%%%%%%%%%%%%%%%%%%%%%%%%%%%%%Chapter Three%%%%%%%%%%%%%%%

\chapter{BPS states and an example}

Although the potential of the dual theory is invariant under T-duality, 
the superspace form of the theory changes under the
transformation. Therefore the   
BPS states and Bogomoln'yi equations of the model
must also change.  
To examine this it is sufficient just to consider the
bosonic sector of \six . Using the standard procedure we express the energy
as a sum of squares plus a topological term
$$\eqalign{
E &= \int dx \, g_{IJ}\left({\dot X}^I{\dot X}^J
+X^{'I}X^{'J}
+{1\over4}m^2s^Is^J \right)\cr
&= \int dx \left(g_{IJ}({\dot X}^I-{1\over2}mk^I)({\dot X}^J-{1\over2}mk^J)
+g_{IJ}(X^{'I}-{1\over2}mu^I)(X^{'J}-{1\over2}mu^J)\right) \cr
&\ \ \ \  + {\cal T} \ . \cr}
\eqn\tena
$$
In \tena\ we have introduced a transparent notation and the `topological' 
term 
$$
{\cal T} = m\int dx \left(k_I {\dot X}^I + u_IX^{'I}\right) \ ,
\eqn\tenb
$$
which is simply a mixture of the Noether charge associated to the symmetry 
generated by $k^I$ and a topological charge arising from the potential.
From \tena\ we can read off a simple set of  Bogomoln'yi equations to be
$$
{\dot X}^I = {1\over 2}m k^I\ , \ \ \ \ \ \ \ \ \ \ 
{X'}^I={1\over 2}mu^I \ .
\eqn\tenc
$$
In the dual model the equations would be simply written in terms of
the  hatted 
fields. One can check from the above formulae that $\cal T$ and 
the $I=i$ components of \tenc\ are unchanged by T-duality. However the $I=0$
components of \tenc\  in the  original and  T-dual model are
$$\eqalign{
{\dot X}^0 &= {1\over 2}m  \ , \ \ \ \ \ \ \ 
{X'}^0 = -{1\over2}mg^{0i}b_{0i}\ ,\cr
{\dot {\hat X^0}} &= 0\ , \ \ \ \ \ \ \ \ \ \   
{\hat X}'^0= {1\over 2}m(g_{00}+g^{ij}b_{0i}b_{0j})  \ .\cr}
\eqn\tend
$$
Thus the BPS states still exist in the T-dual model, but they take on 
a different form. In particular the momentum modes have disappeared 
in the dual model. In the example below we will see that
they have in fact turned into winding modes, in analogy to the
interchange of the string winding and momentum modes under T-duality.

To conclude let  us consider an example given by 
the $(4,4)$ supersymmetric Q-kink sigma model [\AT]
with vanishing torsion and metric
$$
ds^2 = H^{-1}(dX^0 + \omega_idX^i)^2 + H\delta_{ij}dX^i dX^j \ ,
\eqn\exa
$$
where $X^0$ is periodic with period $4\pi$, $i=1,2,3$  and  $\omega_i$ is 
defined by
${1\over2}\epsilon_{ijk}\partial_{j}\omega_{k}  = \partial_i H$. Here 
$H$ is the harmonic function
$$
H = \delta + \sum_{n=1}^{N} {1\over \mid X^i - Y^i_n\mid} \ ,
\eqn\exc
$$
where $\delta = 0,1$ for an ALE or multi-Taub-Nut space respectively, the 
later case describing a KK monopole.
There is a 
unique potential which can be added preserving $(4,4)$ supersymmetry and it
is given as the length of the Killing vector $k=\partial/\partial X^0$;
$$
V = {1\over4}m^2H^{-1} \ .
\eqn\exb
$$
Thus the supersymmetric vacua of this theory are the located at the zeros
of $H^{-1}$, i.e. at the centres $Y^i_n$ of the metric.

If we now dualise along $X^0$ we obtain 
$$\eqalign{
d{\hat s}^2 &= H\left((d\hX^0)^2 + \delta_{ij}d\hX^id\hX^j\right) \ ,\cr
{\hat b}_{0i} &= \omega_i \ ,\cr}
\eqn\exe
$$
which also
admits $(4,4)$ supersymmetry [\CHS] and describes a solitonic 
5-brane. As we showed above the potential \exb\ remains the same, although
now it is given by the length of the non-trivial 1-form 
${\hat u}_I=(1,0,0,0)$. Furthermore one can check that 
$(4,4)$ supersymmetry is preserved by the potential \exb\ in the dual model.

Let us now compare the BPS soliton solutions of these two models. Because our
analysis of these states above assumed only $(1,1)$ supersymmetry it overlooks
the quarternionic structure underlying these models. Let us then briefly
recall the analysis of [\AT] for the model \exa . 
Introducing a triplet ${\bf I}^I_{\ J}$ of complex structures one can 
deduce the richer bound
$$\eqalign{
E&= \int dx \left\{ 
g_{IJ}(X'^I-{1\over2}m({\bf n}\cdot{\bf I})^I_{\ K}k^K)
(X'^J-{1\over2}m({\bf n}\cdot{\bf I})^J_{\ L}k^L) \right. \cr
&\ \ \ \ \ \ \ \ \ \ \ \left.
+g_{IJ}({\dot X}^I-{1\over2}mn_0k^I)({\dot X}^J-{1\over2}mn_0k^J)\right\} 
+m(n_0 Q_0+{\bf n}\cdot{\bf Q}) \ , \cr}
\eqn\exbound
$$
where $(n_0,{\bf n})$ is a unit vector and 
$$
Q_0 = \int dx {\dot X}^Ik_I\ , \ \ \ \ \ 
{\bf Q} = \int dx X'^Ik^J{\bf I}_{IJ}\ .
\eqn\excharge
$$
are the Noether charge and three topological charges.
By choosing $(n_0,{\bf n})$ parallel to $(Q_0,{\bf Q})$ and
setting the squared terms in \exbound\ to
zero we obtain the Bogomoln'yi equations 
$$
{\dot X}^0 ={1\over2}mn_0\ , \ \ \ \ {\dot X}^i = 0 \ , \ \ \ \ 
X'^0=0\ , \ \ \ \ X'^i = {1\over2}mn^iH^{-1}\ ,
\eqn\exbogo
$$
with the energy of the given by  
$m\sqrt{Q_0^2+{\bf Q}\cdot{\bf Q}}$.
Thus one finds the Q-kink
BPS solutions [\AT]\foot{These solutions are only for the case $\delta=0$. For 
$\delta=1$ there also exist solutions but they can not be expressed in 
a closed form.}
$$\eqalign{
X^0 &= X^0_0+{1\over2}mn_0t \ ,\cr
X^i &= {1\over 2}(Y^i_1+Y^i_2) 
+{1\over 2}(Y^i_1-Y^i_2){\rm tanh}\left(
{m\mid{\bf n}\mid\over 16}(x-x_0)\right) \ , \cr}
\eqn\exf
$$
where $X^0_0$ and $x_0$ are arbitrary constants 
and $Y^i_1$ and $Y^i_2$ are any two 
distinct vacua of the potential \exb .

Now consider the T-dual theory. The analysis follows along similar line
except that 
that now we must use $\hu^I$ instead of $k^I$ (the complex structures are
also different in the two models [\Sfetsos]). In this case the charges are
given by (cf. \tenb )
$$
{\hat Q}_0 = \int dx {\hat X'^I} \hu_I\ , \ \ \ \ \ 
{\hat {\bf Q}} = \int dx {\hat X'^I} \hu^J{\hat {\bf I}}_{IJ}\ .
\eqn\exchargetwo
$$
Note that now all the charges are topological.
In order to obtain the correct Bogomoln'yi bound we may write down a similar
expression to \exbound\ 
$$\eqalign{
E&= \int dx \left\{ {\hat g}_{IJ}
({\hat X'^I}-{1\over2}m(n_0\hu^I+{\bf n}\cdot{\hat {\bf I}}^I_{\ K}\hu^K))
({\hat X'^J}-{1\over2}m(n_0\hu^J+{\bf n}\cdot{\hat {\bf I}}^J_{\ L}\hu^L))
\right.\cr
&\ \ \ \ \ \ \ \ \ \ \ \left. 
+{\hat g}_{IJ}{\dot {\hat X^I}}{\dot {\hat X^J}}\right\}   
+m(n_0{\hat Q}_0+{\bf n}\cdot{\hat {\bf Q}}) \ . \cr}
\eqn\exboundtwo
$$
We therefore arrive at the
dual Bogomoln'yi equations 
$$
{\dot {\hat X^I}}= 0\ , \ \ \ \ 
{\hat X'^0} ={1\over2}mn_0H^{-1}\ , \ \ \ \ 
{\hat X'^i} = {1\over2}mn^iH^{-1}\ ,
\eqn\exg
$$
The energy of these solutions given by 
$m\sqrt{{\hat Q}_0^2+{\hat {\bf Q}}\cdot{\hat {\bf Q}}}$ 
and the BPS states of the dual theory are
$$\eqalign{
\hX^0 &= \hX^0_0 + 
{1\over 2}n_0{(\mid Y_1\mid^2-\mid Y_2\mid^2)\over\mid Y_1- Y_2\mid}
+{1\over 2}n_0\mid Y_1-Y_2\mid
{\rm tanh}\left({m\mid{\bf n}\mid\over 16}(x-x_0)\right)
\ , \cr
{\hat X^i} &= {1\over 2}( Y^i_1+ Y^i_2) 
+{1\over 2}( Y^i_1-Y^i_2){\rm tanh}\left(
{m\mid{\bf n}\mid\over 16}(x-x_0)\right) \ . \cr}
\eqn\exdaulbps
$$
Thus the Q-kink momentum modes have become wrapping modes around the compact 
dimension. It is pleasing to see that the notion of T-duality exchanging
momentum and winding modes about the Killing direction extends to the BPS 
states of the worldsheet. Note that the charges, and the hence the masses, 
for the these states are the same before and after T-duality as one would 
expect.
 
\refout

\end